\documentclass[lettersize,journal]{IEEEtran}
\usepackage{amsmath,amsfonts}
\usepackage{algorithmic}
\usepackage{algorithm}
\usepackage{array}
\usepackage[caption=false,font=normalsize,labelfont=sf,textfont=sf]{subfig}
\usepackage{textcomp}
\usepackage{stfloats}
\usepackage{url}
\usepackage{verbatim}
\usepackage{graphicx}
\usepackage{cite}
\usepackage{bm}
\usepackage{makecell} 
\usepackage{xurl}
\usepackage{booktabs}
\usepackage[normalem]{ulem}
\usepackage{amssymb}
\usepackage{xcolor}
\usepackage{fancyhdr}
\usepackage{hyperref}
\usepackage{comment}
\usepackage{xspace}
\usepackage{soul}
\usepackage{upgreek} 
\usepackage{multirow}
\usepackage{wrapfig}
\usepackage{tikz}
\usepackage{anyfontsize}

\def\arch{{SDT}\xspace}

\usepackage{tikz,xcolor,hyperref}

\definecolor{lime}{HTML}{A6CE39}
\DeclareRobustCommand{\orcidicon}{
        \hspace{-2mm}
	\begin{tikzpicture}
	\draw[lime, fill=lime] (0,0) 
	circle [radius=0.16] 
	node[white] {{\fontfamily{qag}\selectfont \tiny ID}};
	\draw[white, fill=white] (-0.0625,0.095) 
	circle [radius=0.007];
	\end{tikzpicture}
	\hspace{-3mm}
}
\foreach \x in {A, ..., Z}{
	\expandafter\xdef\csname orcid\x\endcsname{\noexpand\href{https://orcid.org/\csname orcidauthor\x\endcsname}{\noexpand\orcidicon}}
}

\newcommand{\edit}[1]{{\color{black}{#1}}}
\renewcommand{\sout}[1]{}

\newcommand{\ballnumber}[1]{\tikz[baseline=(myanchor.base)] \node[circle,fill=.,inner sep=1pt] (myanchor) {\color{-.}\bfseries\footnotesize #1};}

\begin{document}

\title{SDT: Cutting Datacenter Tax Through Simultaneous Data-Delivery Threads}

\author{
\begin{tabular}{ccc}
\begin{tabular}{c}
Amin Mamandipoor\\
University of Kansas\\
\href{mailto:aminm@ku.edu}{\textcolor{blue}{aminm@ku.edu}}
\end{tabular}
&
\begin{tabular}{c}
Huy Dinh Tran\\
University of Kansas\\
\href{mailto:huydinhtran@ku.edu}{\textcolor{blue}{huydinhtran@ku.edu}}
\end{tabular}
&
\begin{tabular}{c}
Mohammad Alian\\
Cornell University\\
\href{mailto:malian@cornell.edu}{\textcolor{blue}{malian@cornell.edu}}
\end{tabular}
\end{tabular}

\thanks{Accepted for publication at IEEE Computer Architecture Letters (CAL)}
}

\markboth{}
{Shell \MakeLowercase{\textit{et al.}}: A Sample Article Using IEEEtran.cls for IEEE Journals}

\maketitle

\begin{abstract}
\label{sec:abstract}

Networking is considered a datacenter tax, and hyperscalers push hard to provide high-performance networking with minimal resource expenditure. To keep up with the ever-increasing network rates, many CPU cycles are spent on the networking tax. We make a key observation that network processing threads can be simultaneously executed on server CPUs with minimal interference with the application threads. However, utilizing simultaneous multithreading (SMT) to scale the number of network threads with the number of application threads suffers from (1) failing to provide strict tail latency requirements for latency-critical applications, and (2) reducing the number of available hardware threads for application processes, thus contributing to a high datacenter network tax. In this work, we design, implement, and evaluate a chip-multiprocessor (CMP) with specialized Simultaneous Data-delivery Threads (\arch) per physical core. The key insight is that with judicious partitioning at the architectural level, \arch can safely co-run with application processes with guaranteed performance isolation. Our evaluation results, using full-system simulation, show that a 20-core CMP enhanced with \arch reduces the area and power consumption of a baseline 40-core CMP by 47.5\% and 66\%, respectively, while reducing network throughput by less than 10\%.

\end{abstract}


\section{Introduction}
\label{sec:introduction}
\vspace{-4pt}

\begin{wrapfigure}{r}{0.4\linewidth}
\begin{center}
	\centering
	\vspace{-2ex}
		\includegraphics[width=1.0\linewidth]{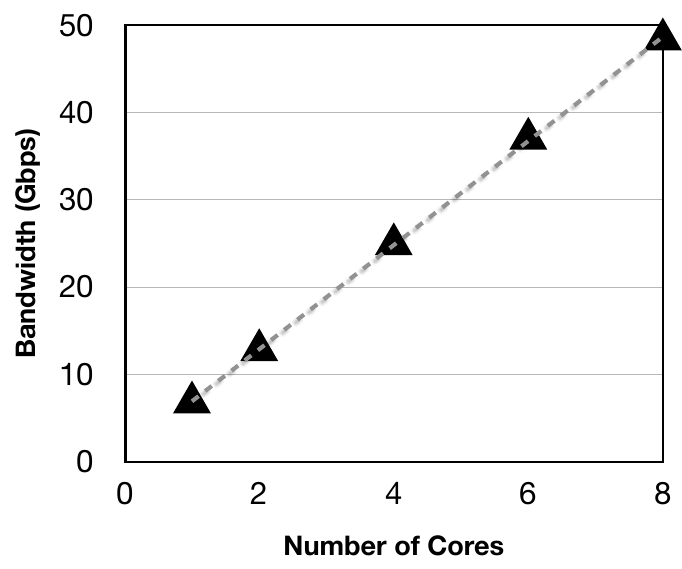}
\caption{Scaling data network delivery bandwidth requires CPU cycles.}
	\label{fig:graph-dpdk-core-scaling}
 \vspace{-23pt}
\end{center}
\end{wrapfigure}

End-host networking bandwidth is increasing exponentially in datacenters. Despite progress in reducing the network software stack overhead and the deployment of userspace networking, delivering data from the NIC to application threads still requires a significant number of CPU cycles. The bandwidth of network data delivery only scales if more CPU cycles are allocated to the userspace networking software. As shown in Figure~\ref{fig:graph-dpdk-core-scaling}, the data delivery bandwidth of DPDK~\cite{dpdk}, the state-of-the-art userspace networking framework, scales linearly when more hardware threads are assigned to the polling mode driver and the simple layer-2 network function.

Another challenge for high-throughput networking is the amplification of on-chip and off-chip data movement during data delivery from the NIC to the application threads. Direct cache access technologies~\cite{ddio_intel_nodate} are developed to minimize off-chip data movement for data delivery from I/O devices. However, while I/O data placement in shared caches can be effective in reducing off-chip data movement, it does not address on-chip data movement. Previous works aimed at reducing both on-chip and off-chip data movement for data delivery from the NIC to CPU, but they only considered run-to-completion network functions~\cite{alian_idio_2022}. These works fall short in addressing the complexity of the network data delivery path, which starts at the NIC, passes through the physical core running the network thread, and finally reaches the physical core executing the application logic.

This work sets out to provide free CPU cycles for delivering network data from the NIC all the way to the core pipelines with minimal on-chip data movement. To achieve this, we introduce the concept of Simultaneous Data-delivery Threads (\arch), where each core pipeline in a Chip Multi-Processor (CMP) adds a minimalist hardware thread that is exposed to the OS where it can be used for running \textit{data-delivery threads}, either in kernel space or userspace networking stacks. The key insight is that the data-delivery threads --- that are responsible for polling NIC, running interrupt handlers, encapsulating/decapsulating headers to/from the payload, enqueuing pointers, etc. --- sporadically utilize a fraction of the hardware resources in a beefy out-of-order superscalar pipeline. \arch leverages this insight to add a minimalist simultaneous hardware thread to each core and judiciously partition the physical resources between the data delivery thread and other hardware threads, minimizing interference while delivering high performance for both the data delivery thread and co-running application threads.

\vspace{-1pt}

\section{Background and Motivation}
\label{sec:background}
\IEEEpubidadjcol
\vspace{-4pt}
\noindent\textbf{Data Delivery vs. Data Processing. } 
We distinguish network \textit{data delivery} functionality from \textit{data processing} functionality. In essence, data delivery includes receiving packets from the datacenter network, transferring them to shared NIC-CPU buffers (through a DMA engine), notifying the CPU of the new packet delivery, and finally executing a \textit{data delivery function} on the CPU that fetches received packet headers, decapsulate them, and copy the data or pass a reference of the received payloads to a \textit{data processing function}. 

Based on this classification, we can segregate any distributed application into two phases: data delivery and data processing. Depending on the physical core and the timing of execution of the \textit{data processing function} after the completion of the data delivery, we can classify network applications into \textit{run-to-completion} and \textit{pipeline} categories. For simplicity, we consider a userspace networking stack, but the same classification can be applied to kernel space networking. 

A \textit{run-to-completion} application implements a blocking loop on the data delivery function and immediately calls the data processing function with pointers to the newly received payloads as input arguments. 
In a \textit{pipeline} application, data delivery and data processing functions run on separate threads and communicate synchronously or asynchronously. The data delivery and data processing threads can timeshare a single core or run in parallel on two different cores.
Although executing the functions on separate cores can enable data delivery and data processing to be performed in parallel, such a pipeline leads to frequent inter-core data movement.

\begin{figure}[!t]
    \centering
    \subfloat[\footnotesize Run-to-completion]{
    \includegraphics[height=0.285\linewidth]{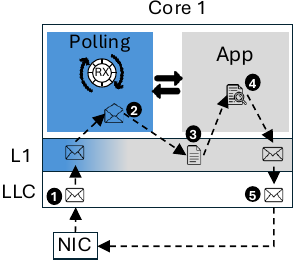}
        \label{fig:runtocompletion}}
    \hfill
    \subfloat[\footnotesize Pipeline]{
    \includegraphics[height=0.285\linewidth]{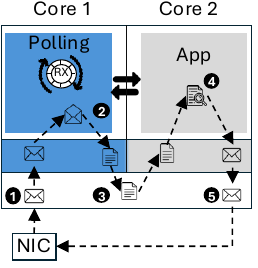}
        \label{fig:pipeline}
    }
    \hfill
    \subfloat[\footnotesize Performance]{
    \includegraphics[height=0.285\linewidth]{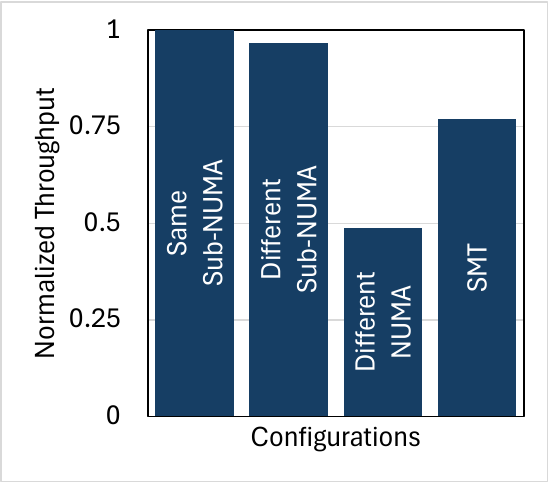}
        \label{fig:iperf-numa}
    }
    \caption{Overview of (a) run-to-completion and (b) pipeline applications datapath, (c) iperf's performance for different NUMA settings.}
    \vspace{-20pt}
    \label{fig:rtc-pipe-numa-graphs}
\end{figure}

Figure~\ref{fig:runtocompletion} and~\ref{fig:pipeline} illustrate the data delivery path of run-to-completion and pipeline applications, respectively. After packets are received from the NIC (\ballnumber{1}), the core running the data delivery thread fetches the packet into its L1 cache for processing (\ballnumber{2}). Once the packet is processed by the data delivery thread, the payload needs to be passed to the data processing thread (\ballnumber{3}). In a run-to-completion implementation, this happens through a simple function call. However, in a pipeline implementation, the payload is passed through an inter-process communication scheme to the data processing thread. The data processing thread then completes the execution of the application (\ballnumber{4}) and sends a response back (\ballnumber{5}).

\noindent\textbf{A Case for Simultaneous Data Delivery Threads.}
One of the key benefits of run-to-completion is eliminating the overhead of inter-core communication between data delivery and data processing functions. However, run-to-completion is very restrictive and not scalable, and except for very simple network functions, a pipeline programming style is used. 

To understand the importance of minimizing data movement between the data delivery and data processing threads, we set up an experiment where we pin the data delivery and data processing threads to different cores with varying distances. We compare the performance of pinning the data delivery and data processing threads to the same \edit{physical core but different hardware threads, same} Sub-NUMA Clusters (SNCs)\footnote{Sub-NUMA Clustering (SNC) is a feature on some Intel processors that splits the processor's memory, cache, and cores into multiple NUMA domains.}, different SNCs, and different NUMA nodes. As shown in Figure~\ref{fig:iperf-numa}, moving the data delivery and data processing threads to different SNCs or NUMA nodes results in a performance loss of 13\% and 37.8\% in the end-to-end iperf performance, respectively. \edit{Sharing the core between data delivery and processing threads, while beneficial in minimizing data movement overhead between the threads, results in a 24\% lower throughput compared to using two physically adjacent cores. This reduction is due to the static partitioning of core pipeline resources and the contention they introduce.} 

In this work, we aim to provide an architecture that offers the performance benefit of run-to-completion but the flexibility of pipeline implementation. We introduce Simultaneous Data Delivery Threads (\arch), where we co-locate data delivery and data processing threads on the same physical core and judiciously partition the physical resources between them to deliver the performance of a run-to-completion running application with significantly fewer physical resources. 

The insight behind designing SDT is that data delivery threads can be executed simultaneously with data processing threads with minimal interference. In the next section, we conduct a design space exploration to identify the minimal resources required for the data delivery thread to keep pace with the processing rate of a data processing thread, which consumes network data as soon as it is received.

\section{Demystifying Network Data Delivery}
\label{sec:demystifying-datadelivery}
\vspace{-4pt}

In this section, we profile the network data delivery path at the microarchitectural level and present several key takeaways that inform the SDT architecture.

\noindent\textbf{Experimental Methodology. } We use DPDK l2fwd as a representative data delivery function, as it has a minimal data processing function that performs a simple MAC address swap. \edit{We consider more complex protocol processing code to be part of the data processing phase.} We run l2fwd on gem5~\cite{lowe-power_gem5_2020}, utilizing a hardware load generator to stress l2fwd~\cite{umeike_userspace_2024}.  The key parameters of the simulated node are detailed in Table~\ref{gem5-config-table}. We take this baseline architecture, which loosely models a beefy Intel Xeon Alder Lake CMP, and sweep the size of the key microarchitectural components, and report data delivery throughput and \edit{tail latency}.

Our experimental methodology consists of a two-phase simulation approach. Initially, we warm up the caches for 2 ms. We then switch to a detailed out-of-order CPU mode for an additional 4 ms of simulation time. During this phase, we activate a hardware-based load generator that injects 8 million 64-byte packets at the maximum rate a single core can sustain.

\begin{table}[ht]
\centering
\fontsize{6.8pt}{6.8pt}\selectfont
\setlength{\tabcolsep}{1pt} 
\begin{tabular}{c|c|c}
\hline
Parameters                  & \makecell{Default (Beefy)} & \makecell{Data-Delivery Requirement (Minimalist)} \\ \hline
CPU Type                    & Out-of-Order                  & \textcolor{black}{Out-of-Order}               \\
Superscalar                 & 12 ways                       & \textcolor{black}{3 ways}                \\
IQ/LQ/SQ                    & 194/144/112                   & \textcolor{black}{32/32/32}              \\
Int/FP/Vec register         & 448/256/400                   & \textcolor{black}{92/0/46}                \\
ITLB/DTLB/BTB/ROB           & 64/64/8192/512                & \textcolor{black}{3/10/256/128}     \\
Private L1-iCache           & 32KB (8-way)                  & \textcolor{black}{4KB}                          \\
Private L1-dCache           & 64KB (16-way)                 & \textcolor{black}{16KB}                \\
Private L2-Cache            & 1MB (16-way)                  & \textcolor{black}{512KB}                \\
Shared L3-Cache             & 2MB (16-way)                  & \textcolor{black}{256KB}                \\
DRAM                        & DDR4 2400MHz/4GB              & \textcolor{black}{DDR4 2400MHz/4GB}               \\ \hline
\end{tabular}
\caption{gem5 configuration and data-delivery requirements.}
\vspace{-8pt}
\label{gem5-config-table}
\end{table}

\begin{figure*}[t]
\begin{center}
        \includegraphics[width=1.0\linewidth]{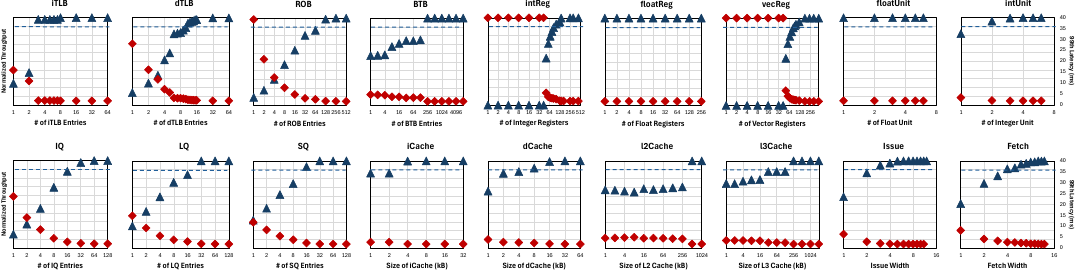}
\caption{Sensitivity of data delivery thread to size of microarchitectural structures. The horizontal line is the 90\% performance watermark.}
\vspace{-25pt}
	\label{fig:graph-dpdk-profiling}
\end{center}
\end{figure*}

\noindent\textbf{Data-Delivery Sensitivity to Microarchitectural Parameters.} 
\edit{As shown in Figure~\ref{fig:graph-dpdk-profiling}, }
the performance of the data delivery thread is fairly insensitive to floating point (FloatReg, and floatUnit) and instruction supply components (iTLB, and iCache). The reason is the lack of floating point operations and the small instruction footprint of the data delivery thread. Although the data delivery performance is sensitive to the size of other microarchitectural structures, it drastically under-utilizes those structures. More specifically, we can reduce the size of key microarchitectural structures by 71\% to 97\% with less than 10\% reduction in the performance of the data delivery thread. The last column of Table~\ref{gem5-config-table} reports a minimal core configuration for a data delivery thread to deliver 90\% of the throughput of the default beefy core.  

\vspace{-1pt}

\section{\arch Architecture}
\label{sec:architecture}
\vspace{-4pt}

In Section~\ref{sec:demystifying-datadelivery}, we discussed the significant opportunity to reduce the resources allocated for data delivery on a CMP. A naive approach to leverage this insight would be to design a heterogeneous CMP with dedicated data delivery cores and data processing cores, similar to big.LITTLE architectures~\cite{kumar_single-isa_2003}. However, such a heterogeneous architecture incurs data-movement overhead between the private caches of the data-delivery and data-processing cores (\S\ref{sec:background}). Additionally, when network activity is low, the data-delivery cores remain underutilized. In this work, we present SDT,
which enhances regular cores on CMPs with a simultaneous multi-threading capability that dynamically partitions core resources between the data delivery and data processing threads based on network load. \arch minimizes data movement by co-locating data delivery and data processing threads on the same physical core and private cache hierarchy.

While traditional SMT implementations
often partition on-chip components symmetrically between threads~\cite{secsmt}, this approach is suboptimal for data delivery tasks, which typically require fewer resources compared to
data processing threads. When running alongside a low compute-intensity application, \arch allocates more resources to fulfill network requirements. Conversely, when paired with a high compute-intensity application, resources are prioritized for the data processing thread.

To accommodate varying workload scenarios, we propose a lightweight mechanism that dynamically adapts resource allocation between data delivery and data processing threads based on real-time demands. We partition resources asymmetrically based on each thread's actual requirements. 
The key insight driving our approach is the ability to significantly reduce the size of key microarchitectural structures for \arch while maintaining near-optimal performance in data delivery to the core.

We introduce several asymmetrically partitioned configurations to leverage the distinct behaviors of data delivery and data processing threads. These configurations are enabled through a software daemon controller which dynamically adjusts resource allocation at runtime. 
Below, we detail each component of our proposed mechanism:

\noindent\textbf{Microarchitecture.} Our mechanism requires minimal modifications to a baseline SMT core that typically partitions its pipeline components equally among co-running threads. We extend this model by providing three additional configurations with asymmetrically sized partitions. In total, each component supports four configurations, allowing for more flexible resource allocation.

Resource allocation varies based on computational intensity. In the \textbf{Baseline} scenario, resources are equally partitioned between threads. For \textbf{High Computational Intensity}, \arch receives 10\% of resources, with the remainder allocated to data processing. In \textbf{Medium Intensity} scenarios, \arch's share increases to 20\%. For \textbf{Low Intensity}, \arch utilizes 40\% of resources, with the rest dedicated to data processing.

Our asymmetric configurations are added at processor design time, building upon existing SMT partitioning mechanisms. For each thread, we utilize two registers per partitioned structure \cite{margaritov_stretch_2019}: a \textit{limit register} and a \textit{usage register}. The limit register defines the maximum number of entries a thread can occupy, while the usage register tracks the current allocation. Control logic checks each cycle to ensure the usage doesn't exceed the limit, blocking further allocation when they are equal.
To enable our asymmetric partitioning scheme, we make the limit registers programmable which allows us to dynamically adjust the maximal occupancy for key structures based on the selected configuration. Our approach requires one register pair per structure, resulting in negligible hardware overhead.
Importantly, this implementation does not necessitate complex control logic beyond what is already present in baseline cores supporting equal partitioning.
\noindent\textbf{Software Daemon.}
To complement our hardware-level resource management, we developed the \arch daemon, a software component that controls the resource partitioning between data delivery thread and main thread by periodically sampling the system performance and adjusting the resource partitioning ratios. The daemon periodically monitors system performance and issues resource partitioning commands. To enable efficient communication between the \arch daemon and hardware structures, we extended the ISA with a custom Store Resource Partition (\texttt{\textit{STRP}}) instruction. This instruction conveys partitioning policies to specific microarchitectural components.

The \arch daemon allows for flexible implementation of different heuristics based on hardware and software telemetry to control the partitioning between SDT and data processing threads. In our work, we demonstrate that a simple heuristic based on network load is sufficient to deliver near-optimal performance.

\noindent\textbf{Re-partitioning Overhead.} The re-partitioning frequency of \arch daemon is set at 1 ms. To prevent data conflicts and ensure data integrity --- given that existing entries may become invalid due to changes in resource ownership among threads --- it is essential to adopt an effective pipeline management strategy. We consider two primary approaches: pipeline flush and pipeline drain.

The pipeline flush method invalidates all in-flight instructions and immediately resumes execution with the new partitioning scheme. This approach facilitates a rapid transition, potentially refilling the ROB and pipeline entries within a few hundreds of cycles, while minimizing overall re-partitioning penalties by bypassing pending instruction completions. In contrast, the pipeline drain method halts the fetching of new instructions, allowing in-flight instructions to complete before applying the new partitioning. While this method maintains instruction continuity, it incurs higher overhead, especially when pending instructions create data flow dependencies on load instructions that miss in on-chip caches. In our simulations, the pipeline flush and drain options incur approximately 200 cycles and 1400 cycles of overhead, respectively. \arch uses the pipeline flush option due to its lower overhead for re-partitioning. In the worst-case scenario, where the software daemon re-partitions every 1 ms, the performance overhead introduced by pipeline flushes is less than 0.001\%.

\vspace{-1pt}

\section{Evaluation}
\label{sec:evaluation}
\vspace{-4pt}

\noindent\textbf{Methodology.} We use gem5's full system mode to evaluate \arch design outlined in Section~\ref{sec:architecture}. The \arch software daemon partitions the following key components based on network load: IQ, LQ, SQ, BTB, ROB, intReg, floatReg, and vecReg. We implemented a DPDK-based micro-benchmark that models a generic two-phase network application in pipeline mode: a data delivery thread that operates similarly to the l2fwd application, and a data processing thread that can be configured with different computational intensities (i.e., the number of CPU cycles per network byte received). We experiment with three levels of computational intensities: low, medium, and high, which correspond to 9Gbps, 4Gbps, and 500Mbps of network load per core. The baseline simulated node is configured with the parameters shown in Table~\ref{gem5-config-table}.

\begin{figure}
    \centering
\begin{center}
	\centering
	\vspace{-3ex}
		\includegraphics[width=0.8\linewidth]{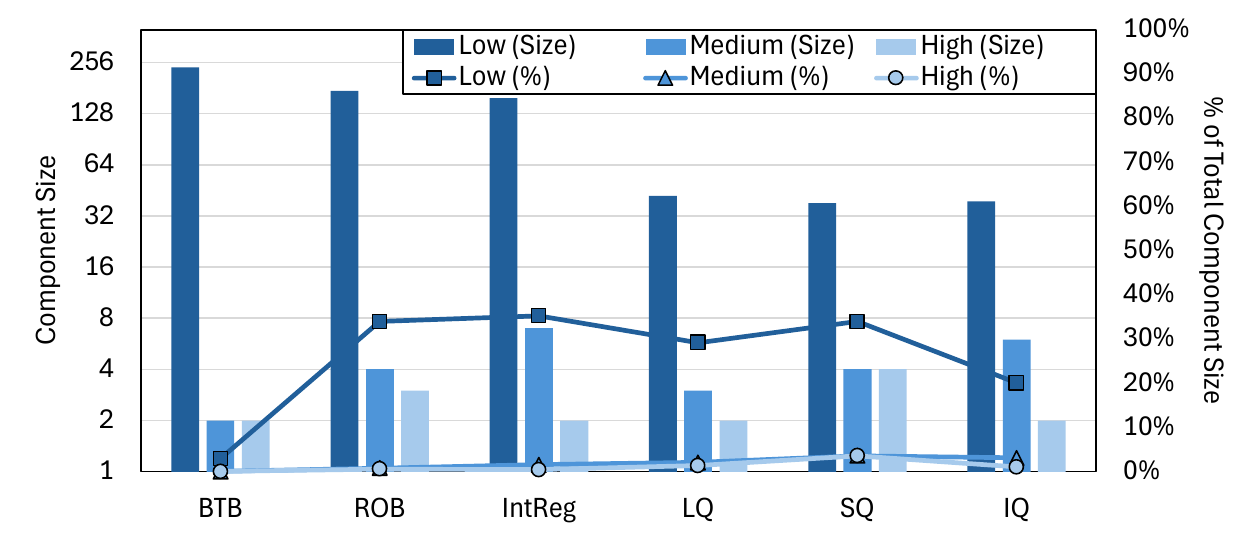}
\caption{ \arch requirements for different levels of compute-intensity applications.}
\vspace{-22pt}
	\label{fig:intensity}
\end{center}
\end{figure}

\noindent\textbf{Results.}
Figure~\ref{fig:intensity} shows the partitioning between the data delivery thread (running on \arch) and the data processing thread (running alongside \arch on the same physical core) to maintain at least 90\% of the performance of a beefy core with 2$\times$ more resources, evenly partitioned between the threads. As illustrated, as the compute intensity of the data processing thread increases, the data delivery rate decreases, and the core pipeline resources naturally shift from \arch to the data processing thread.

Across different compute intensity configurations, \arch is assigned only between 3.6\%--35.3\% of the total pipeline resources across various microarchitectural components. 
Using McPAT~\cite{li_mcpat_nodate}, we compared \arch with the baseline, where each thread hugs a full beefy core, \arch achieves 47.5\% area savings and 66\% power savings for a chip multiprocessor with 20 cores, as configured in Table~\ref{gem5-config-table}.

\section{Conclusion}
\label{sec:conclusion}
\vspace{-4pt}

This work introduces a novel CMP architecture where each core is enhanced with Simultaneous Data-delivery Threads (\arch) to reduce the CPU cycles spent on delivering data from the network to the processing cores. \arch introduces specialized hardware threads that utilize a small fraction of the core pipeline resources without interfering with the main hardware threads executing on the core. By dynamically partitioning physical resources between data delivery and processing threads, \arch maintains 90\% of the baseline CMP performance, where an entire physical core is dedicated to data delivery threads. \arch achieves 47.5\% area savings and 66\% power savings for a 20-core CMP. These results demonstrate the potential of specialized simultaneous hardware threads in the design of next-generation cloud-native CMPs.

\section*{Acknowledgments} This work was supported in part by NSF grant numbers 2239020 and 2311891 and by ACE, one of the seven centers in JUMP 2.0, a Semiconductor Research Corporation (SRC) program sponsored by DARPA. Any opinions, findings, conclusions, and recommendations expressed in this material are those of the authors and do not necessarily reflect those of the sponsors. Amin Mamandipoor and Huy Dinh Tran contributed equally to this work.

\bibliographystyle{IEEEtran}
\vspace{-4pt}
\bibliography{main}

\end{document}